%% file: bose_gas.tex
\pdfoutput=1
\documentclass[aps,eqsecnum,amsmath,twocolumn,notitlepage]{revtex4-1}
\usepackage{graphicx}
\usepackage{amsmath}
\usepackage{amsfonts}
\usepackage{bbold,dsfont}
\usepackage{microtype}
\usepackage{booktabs}

\newcommand{\beq}{\begin{equation}}
\newcommand{\eeq}{\end{equation}}
\newcommand{\bea}{\begin{eqnarray}}
\newcommand{\eea}{\end{eqnarray}}
\def\beqs#1\eeqs{\beq\begin{split} #1 \end{split}\eeq}

\def\dd#1#2{\frac{d #1}{d #2}}
\def\pd#1#2{\frac{\partial #1}{\partial #2}}
\def\av#1{ \left\langle #1 \right\rangle }
\let\Re\relax
\DeclareMathOperator\Re {Re}
\def\Im {\mathop{\hbox{Im}}}
\def\Tr {\mathop{\hbox{Tr}}}

\long\def\comment#1{}

\usepackage[colorlinks=true,backref=false, linktocpage=true,
citecolor=blue,urlcolor=blue,linkcolor=blue,pdfpagemode=UseOutlines]{hyperref}

\hypersetup{%
  bookmarksnumbered=true,
  pdftitle = {},
  pdfsubject = {},
  pdfauthor = {},
  pdfkeywords = {}
}

\begin{document}
\title{A study of symmetry breaking in a relativistic Bose gas using the contraction algorithm}
\author{Andrei Alexandru}
\affiliation{Department of Physics, George Washington University,
Washington, DC 20052}
\affiliation{Department of Physics,
University of Maryland, College Park, MD 20742}
\author{G\"ok\c ce Ba\c sar}
\author{Paulo Bedaque}
\author{Gregory W. Ridgway}
\author{Neill C. Warrington}
\affiliation{Department of Physics,
University of Maryland, College Park, MD 20742}
\date{\today}

%\date{}%
%\dedicatory{}%
%\commby{}%
% ----------------------------------------------------------------

\begin{abstract}
A relativistic Bose gas at finite density suffers from a sign problem that makes direct numerical
simulations not feasible. One possible solution to the sign problem is to re-express the path integral 
%by embedding the original integration contour in a complex space and deforming the integration manifold
%to one with better numerical properties. 
in terms of Lefschetz thimbles.
Using this approach we study the relativistic Bose gas both
in the symmetric phase (low-density) and the spontaneously broken phase (high-density). In the
high-density phase we break explicitly the symmetry and determine the dependence of the order parameter
on the breaking. We study the relative contributions of the dominant and sub-dominant thimbles in this phase. 
We find that the sub-dominant thimble only contributes substantially when the explicit symmetry breaking is small, 
a regime that is dominated by finite volume effects. In the regime relevant for the thermodynamic limit, 
this contribution is negligible. 

%\noindent A relativistic bose gas at finite density is studied using the contraction algorithm, a method for analyzing systems with a sign problem. 
%Integrations over the thimble attached to the global minimal saddle are performed in the spontaneously broken phase, and are compared with the exact result, obtained by integrating over the tangent space of the critical point. Comparison between the two show that the global minimal thimble dominates the path integral in this phase.
\end{abstract}

\maketitle

\input intro
\input algo
\input phi4
\input results

\input conclusions

\acknowledgements

A.A. is supported in part by the National Science Foundation CAREER grant PHY-1151648. 
A.A. gratefully acknowledges the hospitality 
of the Physics Department at the University of Maryland where part of this work was 
carried out.
G.B., P.B., G.R and N.C.W.  are supported by U.S. Department of Energy under Contract No. DE-FG02-93ER-40762.

\bibliographystyle{JHEP}
\bibliography{phi4bib} 
%%%%%%%%%%%%%%%%%%%%%%

\end{document}

%% file: intro.tex
\section{Introduction}
%Sign problem.
%Thimble approach.
%More general manifolds of integration.
%Relativistic Bose gas and spontaneous symmetry breaking.
%Goal of the paper: test the ideas and algorithms related to Monte Carlo integration over manifolds in the complexified space. Probe the contribution of other thimbles.

The Monte Carlo method is a powerful tool to study field theories even when perturbation theory and other analytical methods fail. 
%Many interesting quantum field theories can be studied numerically using Monte-Carlo techniques. 
The method is usually applied in
the Euclidean time formalism where the path integral defining the field theory becomes equivalent to a statistical mechanics problem: each field configuration 
is sampled according to a Boltzmann factor $\exp(-S)$, with $S$ being the Euclidean action for the configuration, 
and the observables are expressed as correlations of composite field operators. 
The Monte Carlo method, however, has its limitations.
When the action is not real, as is often the
case in the presence of   a chemical potential, this approach fails. A possible solution is to 
use the real part of the action $S_R$ for importance sampling and combine the imaginary part with the observable, 
that is, to replace ${\cal O}$ with ${\cal O}\exp(-i S_I)$. For theories that exhibit the {\em sign problem}, the 
phase factor fluctuates substantially and the signal-to-noise ratio for $\av{\exp(-i S_I)}$ decreases exponentially 
fast as the volume increases. This renders the stochastic approach unfeasible. 

Recently it was suggested that the sign problem can be solved by re-expressing the path integral using 
Lefschetz thimbles~\cite{Cristoforetti:2012su}.
The idea is to take the original path-integral expressed in terms of $n$ real degrees of freedom and view it as
an integral over the $\mathds{R}^n$ submanifold embedded in $\mathds{C}^n$. Smooth deformations of the
integration domain will not change the value of the integral as long as we do not cross any singularities of the integrand and
we keep the integral finite. While the integral remains the same, the fluctuations of the integrand on
some of these manifolds could be reduced, making this formulation better suited for numerical sampling.
A particular choice, and in a certain sense the optimal one, is the decomposition of the original integration domain in terms of Lefschetz thimbles,
manifolds that have constant $S_I$~\cite{Cristoforetti:2012su}. If the decomposition involves only one
thimble, or if one thimble dominates the path integral, then the sign problem is solved.
This does not seem to be the case for theories in $0+1$ dimensions, even in the continuum 
limit~\cite{Alexandru:2015xva,Fujii:2015bua,Fujii:2015vha}, but there is still a hope that this might
happen in the thermodynamic limit or in the continuum limit for quantum field theories~\cite{Cristoforetti:2012su}.
We note that even if this is not the case, there is a possibility that alternative methods based on
other manifolds might be numerically useful~\cite{Alexandru:2015sua}.

In this paper we study the relativistic Bose gas at finite density using the contraction algorithm. This system 
has a complex action  leading to a sign problem. In the last few years this system was used as a test case 
for different approaches to the sign problem: complex Langevin~\cite{Aarts:2008wh}, dual 
variables~\cite{Gattringer:2012df,Gattringer:2012ap}, mean field~\cite{Aarts:2009hn,Akerlund:2014mea}, density 
of states~\cite{Bongiovanni:2016jdj}, and Lefschetz thimbles~\cite{Cristoforetti:2013wha,Fujii:2013sra}.
Our goal is to understand the interplay between spontaneous symmetry breaking and the contribution
of non-dominant thimbles, to showcase the contraction algorithm~\cite{Alexandru:2015xva} and some of 
the optimizations we developed~\cite{Alexandru:2016lsn}, and to test some ideas about using 
alternative manifolds as integration domains~\cite{Alexandru:2015sua}. In a recent study it was
conjectured that the tangent space at the critical point of the dominant thimble is equivalent
with the integration over the original domain~\cite{Cristoforetti:2013wha}. We confirm this
conjecture numerically by showing that the results of our algorithm match very well the results
from alternative approaches. Using this result as a benchmark we check the relative contribution of
the dominant thimble.

The plan of the paper is the following. In Section~\ref{sec:algo} we review the relevant details
about the thimbles and describe the algorithm we used to perform the integral over the thimble(s). In Section~\ref{sec:phi4} we introduce the
action for the relativistic Bose gas, its discretization and discuss the details relevant
for our implementation. In Section~\ref{sec:results} we present the results of our method and compare it with results from other approaches.
In Section~\ref{sec:conclusions} we draw conclusions and discuss future directions.

\comment{
Many interesting field theories have a sign problem, which stems from the fact that Euclidean action $S$ describing it is not purely real. The problem with $S=S_R+iS_I$ being complex is that Monte Carlo methods assume that $\text{Pr}(\phi) = \frac{1}{Z}\text{e}^{-S(\phi)} \geq 0$ is a positive definite probability density, which is not the case for a complex action. A remedy to the lack of positive-definiteness of $\frac{1}{Z}\text{e}^{-S(\phi)}$ is to complexify field variables, then deform the integration contour to a set of complex manifolds called ``Lefschetz thimbles". Thimbles are surfaces of constant $S_I$, and so the phase of the action can be separated from its magnitude, yielding a positive definite probability density $\text{Pr}(\phi) \propto \text{e}^{-S_R(\phi)}$. Thimbles are not the only permissible integration surfaces, and in fact a family of surfaces interpolating between the real integration plane and the set of thimbles have demonstrated their utility \cite{Alexandru:2015sua}. A relativistic bose gas at finite density is explored using both the interpolating surfaces, as well as thimbles. The goal of this study is twofold: to probe the relativistic bose gas using new technology, and to explore the distribution of weight amongst the thimbles of the theory in the spontaneously broken phase. 
}

%% file: algo.tex
\section{Lefschetz thimbles and\\ the contraction algorithm}
\label{sec:algo}
%Generalities about critical points and thimbles.
%The algorithm (including the jacobian). Rightians and wrongians.
%$T_{flow}=0 \Rightarrow$ tangent spapce, $T_{flow}>0 \Rightarrow$ other manifolds, 
%$T_{flow}=0 \Rightarrow\infty$ thimbles.

Thermal expectation values of observables in a bosonic field theory have the path integral expression
\beq\label{observable}
\av{\cal O} = \frac1Z \int{D\phi\, e^{-S(\phi)} {\cal O}(\phi) }\quad\text{with}\quad Z=\int{D\phi\, e^{-S(\phi)} },
\eeq
where the integration is taken over real fields $\phi$, and $S$ is the Euclidean action. 
In practice, to compute the path integral above, one first regulates the theory by approximating 
spacetime by a lattice of points. The effect of this regularization on the path integral is to change 
the integration domain from the space of field configurations to $\mathds{R}^n$, where $n$ is the 
number of degrees of freedom. When $S$ is real, the observables can be evaluated by a Monte Carlo integration, that is, $\av{\cal O}$ is estimated as
\beq
\av{\cal O} \approx \frac{1}{\cal N}\sum_{a=1}^{\cal N} {\cal{O}}(\phi_a),
\eeq where the field configurations $\phi_a$ are distributed according to the probability distribution
$\text{Pr}(\phi)=\exp[-S(\phi)]/Z$. The Monte Carlo method does not work when the action $S= S_R + i S_I$ 
is complex. A possible solution to this problem is the so-called ``reweighting" procedure which amounts re-expressing 
the original observable as a ratio of observables, which is amenable to a standard Monte Carlo evaluation:
\beqs\label{reweighting}
\av{\cal O} &= \frac1Z \int{D\phi\, e^{-(S_R + iS_I)} \mathcal{O}} \\
&= \frac1{Z/Z_R} \frac1{Z_R}\int D\phi\, e^{-S_R} \left(\text{e}^{-iS_I} \mathcal{O} \right) = \frac{ \av{ e^{-iS_I}\mathcal{O} }_{S_R} }{ \av{e^{-iS_I}}_{S_R} }
\eeqs
where $\av{\cdot}_{S_R}$ denotes an average with respect to the probability distribution 
$\exp(-S_R)/Z_R$ with $Z_R=\int D\phi\, \exp[-S_R(\phi)]$. Unfortunately the denominator in the equation above 
scales as $\exp(-\beta V\Delta f)$, where $\beta$ is the inverse temperature, $V$ is spatial volume, and 
$\Delta f$ is the free energy difference between the system described by actions $S_R$ and $S$. Consequently, 
the denominator goes to zero exponentially fast as the lattice volumes increases, and any reweighted Monte Carlo is 
dominated by statistical errors. 

An elegant, geometric solution to the sign problem was recently proposed in \cite{Cristoforetti:2012su}. 
The strategy is to first complexify the field variables $\phi$, then deform the integration domain from $\mathds{R}^n$ 
to another submanifold of $\mathds{C}^n$. A particularly judicious choice of submanifold is a set of 
``Lefschetz thimbles", which we will denote by $\{\mathcal{T}_{\sigma} | \sigma=1,2...\}$. We first define thimbles, 
then explain their utility in solving the sign problem. For every critical point
$\phi_c$ of the action, defined by
\beq
\left.\pd S\phi\right|_{\phi_c} = 0 \,,
\eeq
there is an associated thimble. The thimble ${\cal T}$ is defined as the set of all points that, when evolved using 
the \emph{downward flow equations}
\beq\label{downward_flow}
\dd{\phi_i}{t} = -\overline{\frac{\partial S}{\partial \phi_i}},
\eeq
converge to $\phi_c$ as $t\rightarrow \infty$. An analogous \emph{upward flow} is defined by flipping the sign on 
the RHS of Eq.~\ref{downward_flow} and an unstable thimble ${\cal K}$ can be defined with respect to this flow. 
Denoting the complexified field variables $\phi_i = \phi_{R,i} + i \phi_{I,i}$, 
decomposing the effect of the downward flow on its real and imaginary parts and utilizing the Cauchy-Riemann conditions, 
it can be seen that any trajectory along the downward flow follows a negative gradient flow of with respect to $S_R$ and a 
Hamiltonian flow with respect to $S_I$:
\beqs \label{SIconserved}
\frac{d\phi_{R,i}}{dt} &= -\frac{\partial S_R}{\partial \phi_{R,i}} = \frac{\partial S_I}{\partial \phi_{I,i}} \,,\\
\frac{d\phi_{I,i}}{dt} &= -\frac{\partial S_R}{\partial \phi_{I,i}} = -\frac{\partial S_I}{\partial \phi_{R,i}}\,.
\eeqs
The decomposition above shows that thimbles are the multi-dimensional generalization of steepest descent/stationary phase paths from asymptotic analysis.  The imaginary part of the action is then constant over a thimble and it is thus advantageous to deform the region of integration from  $\mathds{R}^n$ to thimbles if there is a rapidly oscillating phase in the integral on $\mathds{R}^n$. This procedure solves the sign problem. Typically, there are many critical points, and hence many thimbles. Moreover it is a non-trivial task to determine what combination of thimbles is equivalent to the original domain of integration $\mathds{R}^n$.
 However, a fundamental 
result of Picard-Lefschetz theory shows that any integral over $\mathds{R}^n$ can be decomposed into integrals over 
thimbles~\cite{Vassiliev:2002su,Milnor:1963su}:
\beq\label{thimble_decomp}
\int_{\mathds{R}^n} d\phi\, e^{-S(\phi)} {\cal O}(\phi) = \sum\limits_{\sigma} n_{\sigma}e^{-iS_I(\phi_\sigma)} 
\int_{{\cal T}_\sigma} d\phi\,e^{-S_{R}(\phi)} {\cal O}(\phi)   \,,
\eeq
where the summation runs over the critical points $\phi_\sigma$ and ${\cal T}_\sigma$ are the associated thimbles.
The integers $n_\sigma$ count the intersection points between the original integration contour $\mathds{R}^n$ and the
unstable thimble ${\cal K}_\sigma$. The $n_\sigma$ can be negative: if the flow takes a volume element around one of these intersection points in $\mathds{R}^n$ into a thimble volume element with orientation opposite to that of the thimble, then the intersection point counts negatively toward $n_\sigma$.  Therefore, $n_{\sigma}$ is the number of points that flow from $\mathds{R}^n$ to the 
critical point $\phi_\sigma$ via the upward flow while preserving orientation, minus the number of intersection points that reverse orientation.

It is in general very difficult to find all critical points and the values of the coefficients $n_\sigma$ in the thimble decomposition above. However, assuming none of the $n_\sigma$ are zero, we see from Eq.~\ref{thimble_decomp} that we can estimate the relative importance of each thimble based on the
value of the real part of the action at the critical point, $S_R(\phi_\sigma)$. The critical point with the lowest action, $\phi_{\bar\sigma}$,
is expected to give the largest contribution to the path integral while the subdominant ones are suppressed by a 
factor $\exp(-\Delta S)$ with $\Delta S = S_R(\phi_\sigma)-S_R(\phi_{\bar\sigma})$. 
This estimate, of course, neglects entropy effects and it is valid only to the extent that the semiclassical expansion is qualitatively correct.
When the main thimble dominates the
integral averages can be approximated by
\beq\label{single_thimble}
\av{\cal O}\approx \int_{\mathcal{T}_{\bar\sigma}}{d\phi\,e^{-S_R(\phi)}\mathcal{O}(\phi)} \Big/ 
\int_{\mathcal{T}_{\bar\sigma}}{d\phi\, e^{-S_R(\phi)}}  \,.
\eeq
Notice that the phase fluctuations are almost absent in this case since the imaginary part of the action factors out
for observable averages and the sampling is done using the real part of the action. The only remaining phase
is the {\em residual phase}, that is the phase associated with the volume element $d\phi/|d\phi|$, which in general
varies smoothly over the thimbles and can be easily reweighted~\cite{Mukherjee:2013aga,Fujii:2013sra,Alexandru:2015xva}.

One of the goals of this paper is to apply the ``contraction algorithm"~\cite{Alexandru:2015xva,Alexandru:2015sua} 
for a relativistic bose gas at finite density. We briefly review the algorithm here. A basic ingredient is the 
map generated by the upward flow $\phi_n\to \phi_f(\phi_n)$
\beq\label{eq:upflow}
\dd \phi t = +\overline{\pd S \phi}\quad\text{with}\quad \phi(0) = \phi_n \quad\text{and}\quad \phi_f=\phi(T_\text{flow})\,.
\eeq
This is a map from $\mathds{C}^n$ to $\mathds{C}^n$. Under this map 
the value of $S_R$ at every point increases while $S_I$ remains fixed. For any integration manifold ${\cal M}_n$ with a finite integral
$\int_{{\cal M}_n} d\phi\, \exp[-S(\phi)]$, it can be shown that the value of the integral is unchanged if we replace
the manifold ${\cal M}_n$ with its image through this map, ${\cal M}_f = \phi_f({\cal M}_n)$~\cite{Alexandru:2015sua}.
Take for the moment ${\cal M}_n$ to be the original integration contour $\mathds{R}^n$. We have
\beqs
&\int_{\mathds{R}^n}d\phi\, e^{-S(\phi)} = \int_{{\cal M}_f} d\phi_f\, e^{-S(\phi_f)} \\
&= \int_{\mathds{R}^n} d\phi_n\, \det J
e^{-S(\phi_f(\phi_n))} \quad\text{with}\quad J_{ij} = \pd{(\phi_f)_i}{(\phi_n)_j} \,.
\eeqs
The last step above is derived using a change of variables from $\phi_f$ to $\phi_n$ with $J$ being the Jacobian of the map.
The contraction algorithm samples the integral using the integrand on the RHS of the expression above. The integrand
is not real, so the sampling is done based on the Boltzmann factor $\exp[-S_\text{eff}(\phi_n)]$ with
\beqs
S_\text{eff}(\phi_n) &\equiv \Re [S(\phi_f(\phi_n)) - \log\det J]\\ &= S_R(\phi_f(\phi_n)) - \log|\det J| \,.
\eeqs
The phase $\varphi(\phi_n)\equiv S_I(\phi_f(\phi_n))-\Im\log\det J$ is reweighted as in Eq.~\ref{reweighting}.
When $T_\text{flow}=0$ the integration manifold  is unchanged and the phase of the integrand  fluctuates rapidly since the
original formulation has a sign problem. As $T_\text{flow}$ increases, $S_R$ increases and the sampled points concentrate on ever smaller  regions of the flowed manifold where $S_R$ is small. As these regions are small and $S_I$ is preserved by the flow, the phase fluctuations on these small sampled regions are also small, alleviating the sign problem. In the $T_\text{flow}\rightarrow\infty$ limit, these sampled regions reduce to isolated points, one for each contributing thimble.
%
%In the limit $T_\text{flow}\to\infty$ the sampled points are infinitesimally closed to these intersection points
%and the phase in the pocket corresponding to $\phi_\sigma$ has a fixed contribution
%$S_I(\phi_\sigma)$ and a fluctuating one coming from $\Im\log\det J$. This latter phase is the residual
%phase: the neighborhood of the intersection point is mapped onto the thimble attached
%to $\phi_\sigma$ and the phase of the Jacobian at $\phi_n$ corresponds to the residual phase, $d\phi/|d\phi|$, at the 
%image point $\phi_f$ on the thimble. 
On the flowed manifold, for large $T_\text{flow}$ , the regions with an important statistical weight form isolated pockets with particularly small $S_R$, each one corresponding to a particular contributing thimble.
In this formulation, the full value of the integral is recovered
only when all the pockets are sampled, corresponding to the inclusion of all thimbles appearing in the decomposition.
Algorithms involving incremental changes in the field configurations will only sample the pocket closest
to the starting configuration. This is actually how we employ our algorithm when we are interested in sampling
only the contribution of one thimble. If we want to include the contribution of other thimbles we have to
also make proposals that can take us from one pocket to another. We used this type of large updates previously in
a fermionic model~\cite{Alexandru:2015sua}, and in Section~\ref{sec:results} we will present a similar procedure 
for the model we study in this paper.

%comment

%end comment

%\subsection{Generic Implementation of the Method}

The contraction algorithm is simply a Metropolis algorithm using the variables $\phi_n$ based on the effective action $S_\text{eff}[\phi_f(\phi_n))]$. The proposals have
to be chosen carefully to get a reasonable acceptance rate. A number of optimizations are possible when the starting
manifold ${\cal M}_n$ is not the original integration domain but the tangent manifold to a thimble of interest. This manifold
is a legitimate choice in two cases: either  we are interested in sampling only one thimble in the limit 
$T_\text{flow}\to\infty$~\cite{Alexandru:2015xva}, or  the tangent space to the thimble is equivalent 
to the original integration domain, which is the case for a class of systems. It was conjectured that this is also true for 
the model studied in this paper~\cite{Cristoforetti:2013wha}. We will show evidence supporting this conjecture.

There are two advantages to using the tangent space of a critical point as the starting manifold instead of $\mathds{R}^n$: we can make  
efficient Metropolis proposals~\cite{Alexandru:2015xva} and we can use an estimator to take into account the contribution
of the Jacobian to the effective action~\cite{Alexandru:2016lsn}. The Jacobian can be computed exactly by integrating
the following differential equation along the upward flow path,
\beq
\dd Jt = \overline{H} \overline{J} \,,
\eeq
where $J$ is the Jacobian matrix and $H$ is the Hessian of the action $S$. The initial condition for $J$ is a matrix
that has as columns a set of vectors that span the tangent space at the critical point. We use
$J(0)=( \hat\rho_1, \ldots,\hat\rho_n)$, where $\hat\rho_i$ are the positive ``eigenvectors'' of the Hessian at the critical point,
that is $\overline{H_0\hat\rho_i}=\lambda_i\hat\rho_i$ with $\lambda_i>0$. It can be shown that these vectors span
the tangent space to the thimble ${\cal T}$ and that $i\hat\rho_i$ are negative ``eigenvectors'' that span the tangent
space to the unstable thimble ${\cal K}$. Integrating the equation above is numerically expensive, since every step of 
the integrator involves multiplication of $H$ and $J$ matrices. Fortunately, there is a very good estimator for $\log|\det J|$ built out of the ``eigenvectors'' $\hat\rho_i$~\cite{Alexandru:2016lsn}:
\beq\label{eq:estimator}
W = \int_0^{T_\text{flow}} dt\, \sum_{i=1}^{n}{\hat\rho_i^{\dagger}\overline{H(t)\hat\rho_i}} \,.
\eeq
Since $H$ is sparse, the cost of computing $W$ is ${\cal O}(n)$, which is much cheaper than the ${\cal O}(n^2)$ cost of computing $J$; the savings
are substantial even for small lattices. The algorithm samples configurations with an effective action
$S_\text{eff}'(\phi_n)=S_R(\phi_f(\phi_n))-\Re W$ and the difference $\Delta S = S_\text{eff}(\phi_n)-S_\text{eff}'(\phi_n)$
is included as a reweighting factor. To compute the reweighting factor exactly we need to integrate the equations
for $J$, but this needs to be done only for the small subset of configurations used for measuring observables, which are separated
by  a large number of Metropolis steps.

%comment

%% file: phi4.tex
\section{Relativistic Bose Gas at Finite Density}
\label{sec:phi4}

%Lattice action.
%(Constant field) critical points. Stokes phenomenon and the need for $h$ complex.
%What we know about the thimble decomposition.
%Sign problem in $\mathds{R}^N$ is really bad.

The Euclidean action of a gas of relativistic spinless bosons in three (spatial) dimensions at finite density is 
given by
\beqs\label{act}
S = \int d^4x\,\big[&\partial_0\phi^* \partial_0\phi + \nabla\phi ^*\cdot\nabla\phi + 
(m^2-\mu^2)\lvert\phi\rvert^2 \\&+ \mu\underbrace{(\phi^* \partial_0\phi - \phi\partial_0\phi^*)}_{j_0(x)}  + 
\lambda {\lvert \phi \rvert}^4 \big]
\eeqs
where $\phi\equiv (\phi_1+i\phi_2)/\sqrt2$ is a complex scalar field. The term involving the boson density $j_0(x)$ is 
imaginary and is the source of the sign problem in this model. This action is symmetric under global $U(1)$ transformations
$\phi\to e^{i\alpha}\phi$. For  values of $\mu$ below a critical value of order $ m$  the equilibrium state
is expected to be $U(1)$ symmetric and $\av\phi =0$. For  values of $\mu$ larger than the critical value and for low enough temperatures the $U(1)$ symmetry is expected to be spontaneously broken and
 $\av\phi\not=0$.
 
  In order to study spontaneous symmetry breaking it is necessary to introduce 
an {\it explicit} symmetry breaking term $S \to S - h\int{d^4x \, (\phi_1 + \phi_2)}$. This choice 
breaks the original $U(1)$ symmetry down to the $\mathds{Z}_2$ sub-group given by swapping 
$\phi_{x,1}\leftrightarrow\phi_{x,2}$.

\begin{figure*}[t]
\includegraphics[width=\textwidth]{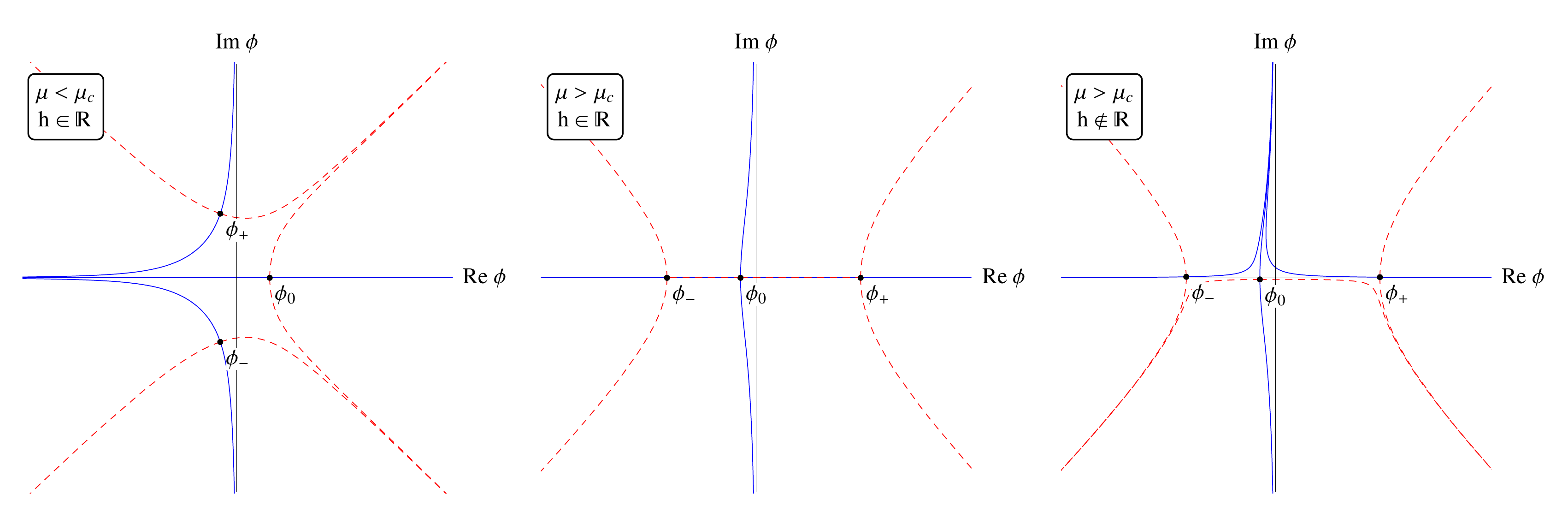}
\caption{Projections in the constant $\phi_1=\phi_2$ subspace for $\mathcal{T}_{0}$, $\mathcal{T}_{+}$ and 
$\mathcal{T}_{-}$, the thimbles attached to $\phi_{0}$, $\phi_{+}$ and $\phi_{-}$. The left and middle graph correspond
to  $h$ real and the one on the right to a value of $h$ that has a small imaginary component. Thimbles $\mathcal{T}_i$ 
are plotted in blue, unstable thimbles $\mathcal{K}_i$ are plotted with dashed, red lines, and critical points are indicated by black 
dots. \label{fig:stokes}}
\end{figure*}

We use  the following discretization of the action:
\beqs\label{discAct}
S =  a^4\sum_x \Bigg[&
\frac{e^{\mu a}\phi^*_{\smash{x+\hat{0}}}-\phi^*_x}{a}\frac{e^{-\mu a}\phi_{\smash{x+\hat{0}}}-\phi_x}{a} 
\\&+ \sum_{\nu=1}^{3}\Big\lvert\frac{\phi_{\smash{x+\hat{\nu}}}-\phi_x}{a}\Big\rvert^2 
+m^2\lvert\phi_x\rvert^2 \\&+ \lambda\lvert\phi_x\rvert^4\ - h(\phi_{x,1} + \phi_{x,2})
 \Bigg]\,.
\eeqs
Setting the lattice spacing to $a=1$ and writing the field in terms of real components, $\phi_1$ and $\phi_2$, we obtain 
\beqs\label{lat_act}
S =  \sum_x \Bigg[&\Big(4+\frac{m^2}{2}\Big)\phi_{x,a}\phi_{x,a}-\sum_{\nu=1}^{3}\phi_{x,a}\phi_{\smash{x+\hat{\nu},a}} 
\\&- \cosh\mu\ \phi_{x,a}\phi_{x+\hat{0},a}  +  i\sinh\mu\ \epsilon_{ab}\phi_{x,a}\phi_{\smash{x+\hat{0},b}}  \\
     &+ \frac{\lambda}{4}  \big(\phi_{x,a}\phi_{x,a})^2 - h (\phi_{x,1}+\phi_{x,2}) \Bigg]
\eeqs
with $\epsilon_{ab}$ the antisymmetric tensor with $\epsilon_{12} = 1$. The derivatives of the action are
\beqs
\pd{S}{\phi_{x,a}} =& (8+m^2)\phi_{x,a}-\sum_{\nu=1}^3 \left( \phi_{x+\hat\nu,a}+\phi_{x-\hat\nu,a} \right) 
\\&- \cosh\mu \left(\phi_{x+\hat0,a}+\phi_{x-\hat0,a} \right) 
\\&+ i\sinh\mu \epsilon_{ac}\left( \phi_{x+\hat0,c}-\phi_{x-\hat0,c} \right) \\
&+\lambda(\phi_{x,c}\phi_{x,c})\phi_{x,a} - h \,, \\
\frac{\partial^2 S}{\partial \phi_{x,a}\partial\phi_{y,b}} =& (8+m^2)\delta_{xy}\delta_{ab}
-\sum_{\nu=1}^3 \left(\delta_{x+\hat\nu,y}+\delta_{x-\hat\nu,y} \right)\delta_{ab} \\
&-\cosh\mu \left(\delta_{x+\hat0,y}+\delta_{x-\hat0,y} \right)\delta_{ab} \\
&+i\sinh\mu \left(\delta_{x+\hat0,y}-\delta_{x-\hat0,y} \right)\epsilon_{ab} \\
&+\lambda \delta_{xy} \left[ (\phi_{x,c}\phi_{x,c})\delta_{ab}+2\phi_{x,a}\phi_{x,b} \right] \,.
\eeqs
It is very difficult to find all critical points of the (latticized) action. However, critical points with small $S_R$ play a special role as their thimbles are more likely to dominate the path integral. Critical points with constant fields $\phi_{xa}=\phi$ are among the ones with smaller values of the action and, at the same time, are easy to find. In fact, 
 the global minimum of the real part of the action restricted to real values of $\phi_{x,1}$ and $\phi_{x,2}$  is a constant field (assuming $h$ real). This can be seen by looking at Eq.~\ref{act} or its discretized version Eq.~\ref{discAct} and noticing that i) the kinetic and gradient term are positive definite and favor constant fields and ii) the term linear in $\mu$ is purely imaginary and does not contribute to the real part of the action. This motivates us to find the constant field critical points.
They are given by $\phi_{x,1}=\phi_{x,2}=\phi$ where $\phi$ is one of the roots of the cubic equation:
\beq\label{eq:cubic}
(2+m^2)\phi -2 \cosh\mu \phi + \lambda |\phi|^2 \phi =h,
\eeq whose three solutions are given, for small $h$, by:
%It is expected that
%the critical points with the lowest action correspond to field configurations that are constant in space and time. 
%For constant fields the gradient vanishes only when $\phi_{x,1}=\phi_{x,2}=\phi_\text{cr}$, with $\phi_\text{cr}$
%satisfying the depressed cubic equation
%\beq
%\phi_\text{cr}^3+\frac{1+m^2/2-\cosh\mu}\lambda \phi_\text{cr}-\frac h{2\lambda} = 0\,.
%\eeq
%which can be solved using Vieta's substitution. The three solutions of this equation expanded around
%$h=0$ are
\beqs
\phi_0 &= -\frac{h}{\hat\mu^2-m^2} + \mathcal{O}(h^2)\,, \\
\phi_\pm &= \pm \sqrt{\frac{\hat\mu^2-m^2}{2\lambda}} + \frac{h}{2}\frac{1}{\hat\mu^2-m^2}+\mathcal{O}(h^2),
\eeqs 
with $\hat\mu^2 = 2\cosh\mu-2$. The  corresponding values of the action at these points are:
\beqs\label{eq:critical_acts}
\frac{S(\phi_0)}{V_4} &= 0+{\cal O}(h^2)\,,\\ 
\frac{S(\phi_\pm)}{V_4}  &= -\frac{1}\lambda \left(\frac{\hat\mu^2-m^2}2 \right)^2
\mp 2h \sqrt{\frac{\hat\mu^2-m^2}{2\lambda}}+{\cal O}(h^2) \,,
\eeqs
where $V_4=(V/a^3)/(aT)$ is the number of lattice sites. 

We now turn to discuss what we know about the thimble decomposition of the original integral over real fields.
For $\mu$ below $\mu_c=\cosh^{-1}(1+m^2/2)$ the critical points
$\phi_\pm$ are imaginary. In this case the thimbles 
${\cal T}_\pm$ associated with $\phi_\pm$ do not contribute to 
the thimble decomposition. This is because the real part of the action in $\mathds{R}^n$ is bounded 
from below by $\Re S(\phi_0)$ which is larger than $\Re S(\phi_\pm)$. Thus no point in the original integration domain can flow 
into $\phi_\pm$ under steepest-ascent/upward-flow defined in Eq.~\ref{eq:upflow}. 
This is illustrated in the the left panel of Fig.~\ref{fig:stokes} which shows a projection of the thimbles ${\cal T}_{0,\pm}$ in the $\phi_1=\phi_2$ plane.  We can see that, in the $\mu<\mu_c$ case, the unstable thimbles ${\cal K}_\pm$ do not intersect the original integration domain, which is the real axis in this figure.
For values of $\mu$ greater than $\mu_c$ the situation changes. 
In  the case  where $\mu>\mu_c$ and  $h$ is real, there is a trajectory of the flow connecting $\phi_0$ and $\phi_+$ ($\phi_-$). Indeed, in the constant field subspace (that is, in the subspace where $\phi_{x,1}=\phi_{x,2}=\phi$)
the gradient 
\beq
\left.\frac{\partial S}{\partial \phi_{x,a}}\right|_{\phi_{x,1}=\phi_{x,2}=\phi}=
(m^2-\hat\mu^2)\phi + 2\lambda \phi^3-h
\eeq 
points along the constant field subspace (all components are equal).  Moreover, the gradient is real if $\phi$ is real. This implies in the existence of flow trajectories lying entirely on the real constant field subspace. Since the downward flow decreases the value of the real part of the action, we conclude that there is a trajectory connecting the real constant fields $\phi_0$ and $\phi_+$ (and another connecting $\phi_0$ to $\phi_-$). Thus, the unstable thimble of $\phi_0$ ($\mathcal{K}_0$) overlaps the thimble of $\phi_+$ ($\mathcal{T}_+$). The existence of these lines is known as a Stokes phenomenon and invalidates the decomposition of the integral into integer linear combinations of thimbles.
%
%Turning to the thimble decomposition, when the symmetry breaking parameter $h$ is real and $\mu>\mu_c$, there is a 
%Stokes line linking $\phi_{+}$ and $\phi_{-}$ (see Fig.~\ref{fig:stokes}). In this case the thimble decomposition is
%ambiguous. 
We bypass this problem by making $h$ slightly complex.
As can be seen from Eq.~\ref{eq:critical_acts} a complex value of $h$ implies different values for the imaginary parts of $S(\phi_+), S(\phi_-)$ and $S(\phi_0)$. As the flows preserves the imaginary part of the action, a complex value of $h$ guarantees that there is no Stokes lines connecting these critical points.
%eliminate this ambiguity by 
%making the symmetry breaking term slightly complex, $h=\lvert h \rvert e^{i\epsilon}$ with $\epsilon \ll 1$. 
The way a complex value of $h$ makes the thimble decomposition well defined is shown visually
 in the center and  right panels of Fig.~\ref{fig:stokes}. 
The figure suggests that the integral over the real line is equivalent to the integral over ${\cal T}_+$ and ${\cal T}_-$ (with the proper orientations), which would imply that $n_+=n_-=1, n_0=0$. 
The Fig.~\ref{fig:stokes} only shows a  projection of the whole space 
but some definite conclusions can be drawn. For instance, the fact that there is a flow line connecting the real plane to $\phi_-$ shows that $\cal K_-$ does intersect $\mathds{R}^n$, although the possibility remains that there are other (non-constant field) points where $\cal K_-$ intersects $\mathds{R}^n$. This is  a strong but not definitive case that $n_-=1$.
The fact that $n_+= 1$ can be
argued even more rigorously.
 The difference in the real part of the action between $\phi_+$ and the global minimum on the real plane is proportional to $\Im(h)$. In the $\Im(h)\rightarrow 0$ limit this region shrinks to a point
\footnote{in some singular cases this region can shrink to a line or other manifold with dimension larger than zero but smaller than $n$. This possibility can be excluded by noticing that, for small $\Im(h)$, the global minimum of $S_R$ on $\mathbb{R}^n$ is very close to $\phi_+$, close enough that the quadratic approximation to the action applies. Within the gaussian approximation the flow equations can be solved analytically and the requirement that the region with $S_R$ smaller than $S_R(\phi_+)$ shrinks to a point can be translated on a condition over the ``eigenvectors'' $\hat\rho_i$. We have verified this condition numerically.} 
showing that, at least in this limit, $n_+=1$. 
%Similarly, the difference of the real part of the action between $\phi_-$ and the global minimum on the real plane is proportional to $\Re(h)$ and vanishes in the $\Re(h)\rightarrow 0$ limit. However, in the $\Re(h)\rightarrow 0$ limit a annular region of points in the real plane have a real part of the action smaller than $\Re(S(\phi_-))$ and it is not obvious that $n_-=1$. 
To the extent that the constant field projection can be trusted we also have $n_0=0$.
Unfortunately, the picture away from the constant field subspace is much harder to analyze and there is the possibility that ${\cal T}_0$, or even that thimbles corresponding to other, non-constant field critical points, may also contribute.
Even if the unstable thimble ${\cal K}_0$ were
to intersect the real integration domain at a point not included in the projection in Fig.~\ref{fig:stokes}, and therefore contribute to the thimble decomposition of the original integral, its
contribution is expected to be much smaller than the one from ${\cal T}_-$. 
In fact, the leading order estimate (in the semiclassical expansion) for the relative contribution of two thimbles, say  ${\cal T}_0$  and ${\cal T}_+$ , is given by the
ratio of Boltzmann factors $\exp(-[S(\phi_0)-S(\phi_+)])$. Since the difference in the action at $\phi_0$ and $\phi_+$ is approximately $(\hat\mu^2-m^2)^2/4\lambda$, the contribution of ${\cal T}_0$, if not identically zero, is exponentially small as $V_4\gg1$ (or $\hat\mu \gg m$ or $\lambda \ll 1$).

The last topic to discuss in this section is the evaluation of the estimator in Eq.~\ref{eq:estimator}. This is done
by integrating the differential equation
\beq
\dd Wt = {\Tr{}' H}\quad\text{with}\quad \Tr{}' H \equiv \sum_i \hat\rho_i^\dagger \overline{H(\phi(t)) \hat\rho_i}
\eeq
together with the upward flow in Eq.~\ref{eq:upflow}. These differential equations are integrated using a Cash-Karp integrator,
an adaptive step-size integrator of order ${\cal O}(\Delta t^5)$~\cite{Cash:1990:VOR:79505.79507}. For every integrator step 
we need to evaluate $\Tr{}' H(\phi(t))$, which requires the positive ``eigenvectors'' $\hat\rho_i$ for the Hessian at the critical 
point $\phi_\text{cr}$
\beq\label{eq:heigensys}
\overline{H(\phi_\text{cr}) \hat\rho_i} = \lambda_i \hat\rho_i\,,\quad\text{with}\quad \lambda_i>0 \,,
\eeq
where
\beq
H(\phi)_{x,a;y,b} = \frac{\partial^2S}{\phi_{x,a}\phi_{y,b}}\,,\quad\text{and}\quad \hat\rho_i^\dagger \hat\rho_j=\delta_{ij}
 \,.
\eeq 
We stress that the estimator involves only the ``eigenvectors'' of the Hessian at the critical point and not the ones
of $H(\phi)$. This estimator is exact when the action is quadratic and it tracks well the true value of the
Jacobian even when the thimble is curved~\cite{Alexandru:2016lsn}. 
We compute the ``eigenvectors'' before starting the Monte-Carlo simulation. This step is computationally costly but needs to be performed only once. 
To evaluate the estimator, we separate the $\phi$-dependent part from the Hessian as:
\beqs
H(\phi)_{x,a;y,b} =& H(\phi_\text{cr})_{x,a;y,b} + \lambda \delta_{xy} 
\big[(\phi_{x,c}\phi_{x,c})\delta_{ab}\\&+2\phi_{x,a}\phi_{x,b}-(\phi\to\phi_\text{cr})\big] \,.
\eeqs
We have then
\beqs
\Tr{}' H(\phi) &= \Tr{}' H(\phi_\text{cr}) + \overline{\Delta(\phi)} - \overline{\Delta(\phi_\text{cr})}\\
&= \left( \sum_i \lambda_i - \overline{\Delta(\phi_\text{cr})} \right) + \overline{\Delta(\phi)}\,,
\eeqs
where we used the fact that $\Tr{}' H(\phi_\text{cr}) = \sum_i \lambda_i$ and the definitions 
\beqs
\Delta(\phi) &\equiv \sum_x \sum_{abc} {\cal R}_{x,ab} \left[ \phi_{x,c}\phi_{x,c} \delta_{ab} + 2\phi_{x,a}\phi_{x,b}\right]\,,\\
{\cal R}_{x,ab} &\equiv \sum_i (\hat\rho_i)_{x,a} (\hat\rho_i)_{x,b} \,.
\eeqs
Note that $\Tr{}' H(\phi_\text{cr})$, $\Delta(\phi_\text{cr})$, and ${\cal R}$ are computed once at the beginning of the
simulation and we only need to evaluate $\Delta(\phi)$ along the integration path. This step has computational 
cost of order ${\cal O}(n)$.

%%%%%%%%%

%% file: results.tex
\begin{figure*}[th]
\includegraphics[width=0.463\textwidth]{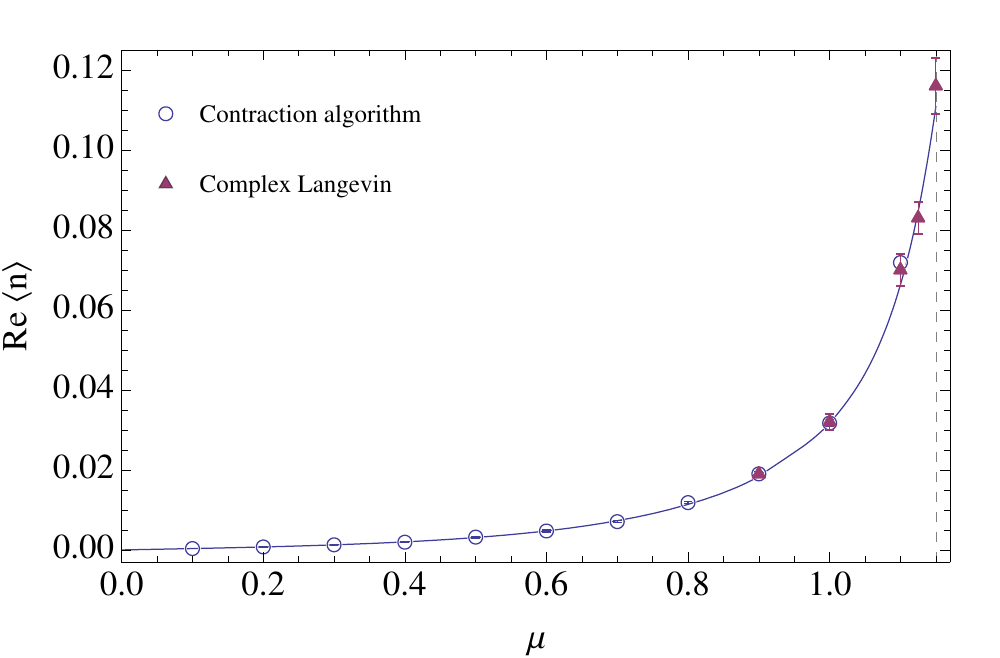}
\includegraphics[width=0.45\textwidth]{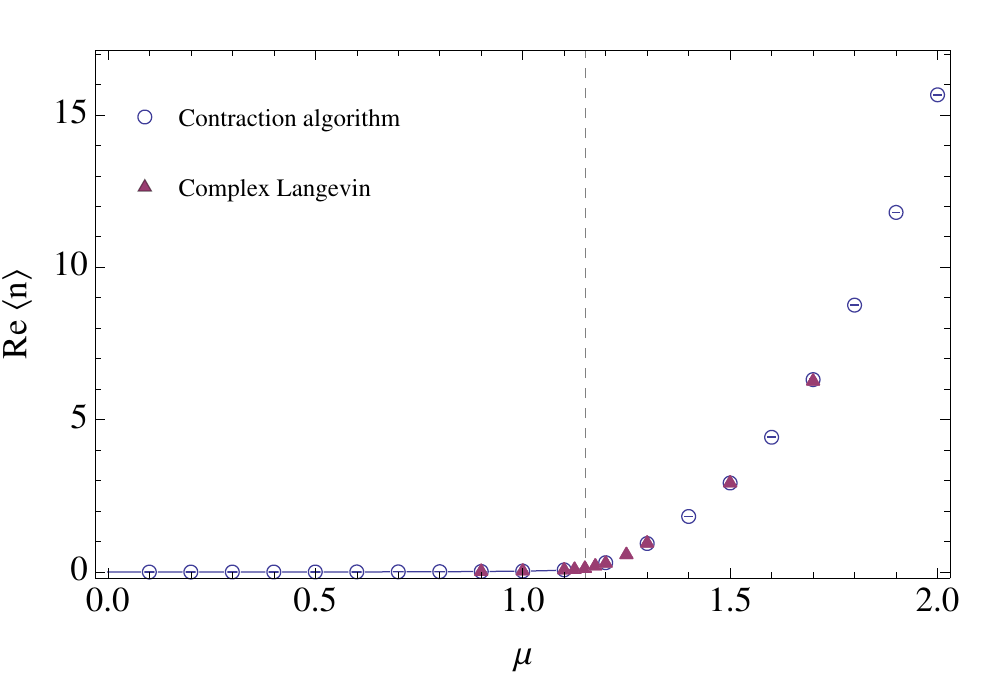}
\caption{Dependence of the charge $\Re\av n$ on the chemical potential on a $4^4$ lattice with parameters $m=\lambda=1.0$
and $h=10^{-3}+i 10^{-4}$. 
In the left panel we plot only the data below $\mu_c^\text{mean}\approx1.15$, the mean field estimate for the transition
to the symmetry broken phase~\cite{Aarts:2009hn}. Complex Langevin data is taken from~\cite{Aarts:2008wh} and the curve 
in the left panel from a mean field calculation that is known to agree very well with exact results 
results~\cite{Aarts:2009hn}.
\label{fig:flow0}}
\end{figure*} 

\section{Results}
\label{sec:results}
%$\langle n \rangle \times h, \langle \phi \rangle \times h$, comparison with literature. Absence of runaway solutions. Tangent plane is probably in the right homology class.
%$\langle greg \rangle\approx 1$ but it'll be smaller at large volumes: motivation for flowing.

In this section we present the results of our simulations. We discuss first the exact results, based on simulations
on the thimble tangent plane at the $\phi_+$ critical point, both in the symmetric and the broken phase. 
We then focus on the broken phase and analyze the dependence of the order parameter as a 
function of the breaking parameter $h$ to determine the regime where finite volume effects are not dominant. 
After that we report the results of single thimble simulations on the dominant thimble, ${\cal T}_+$, and compare them with exact results. Finally, we present a method to carry out simulations on two thimbles and study the relative weight
of ${\cal T}_-$ to ${\cal T}_+$ as a function of $h$.

As we discussed in Section~\ref{sec:algo}, we can use the contraction algorithm to get exact results by flowing
the original domain of integration $\mathds{R}^n$. However, if the starting
manifold is the tangent space at the critical point on a thimble then shorter flow times are required to reduce the phase fluctuations and we can use better proposals for the Metropolis
algorithm and fast Jacobian estimators. 
This choice is justified if we can show that this manifold is equivalent
with the original integration domain. Cristoforetti~{\it et~al}~\cite{Cristoforetti:2013wha} conjectured this to be 
the case for the relativistic Bose gas: they argue that the original integration domain is equivalent to the tangent 
space for the thimble corresponding to the global minimum of $S_R$ over $\mathds{R}^n$. They use real values of $h$ 
and the global minimum corresponding to the constant field critical point $\phi_0$ for $\mu<\mu_c$ and $\phi_+$ for 
$\mu\ge\mu_c$. Note that the transition from $\phi_0$ to $\phi_+$ is smooth since they are equal for $\mu=\mu_c$.
The manifolds $T_0$ and $T_+$ are the tangent manifolds to the thimbles ${\cal T}_0$ and ${\cal T}_+$ at
the critical points $\phi_0$ and $\phi_+$ respectively. Note that when using a slightly complex $h$, as we use, the
points $\phi_0$ and $\phi_+$ have small complex components, proportional to $\Im h$. The tangent spaces $T_0$ and $T_+$
are not parallel to $\mathds{R}^n$, even in the limit $h\to0$, so an analytical proof of the conjectured equivalence
cannot be easily established.

We carried out simulations using $T_\text{flow}=0$ using as the starting manifold $T_0$ for $\mu<\mu_c$ and $T_+$ for
$\mu\ge\mu_c$. To make sure that we stay on the manifold we determine the ``eigenvectors'' $\hat\rho_i$ and parametrize
each point on the manifold as $\phi=\phi_\text{cr}+\sum_i c_i \hat\rho_i$, where $c_i$ are real coefficients. We use
$c_i$ to represent the points in the tangent space, and the updates are done one coefficient at a time with a step
drawn from an uniform probability distribution in the interval $[-\theta,\theta]$ where $\theta=\Delta/\sqrt{\lambda_i}$.
This type of proposal scales the step size in each direction so that the increase in the action due to displacements in each direction are comparable, at least in the
region where the action is well approximated by its quadratic approximation around $\phi_\text{cr}$. We tune $\Delta$
to get a good acceptance rate. In Fig.~\ref{fig:flow0} we show the results for the charge
\beqs
\av{n}&=\frac1{V_4}\pd{\log Z}\mu \\&= \frac1{V_4}\sum_x (\delta_{ab}\sinh\mu-i\epsilon_{ab}\cosh\mu)\phi_{x,a}\phi_{\smash{x+\hat0},b} \,.
\eeqs
In these simulations we set $h=10^{-3}+i 10^{-4}$ and $\Delta=3.0$ so that the acceptance rate was close to 50\%.
For each value of $\mu$ we evaluated $5\times10^6$ updates and performed measurements on configurations separated
by $1000$ updates. We compare our results with the results from Complex Langevin simulations~\cite{Aarts:2008wh}, which are known
to agree with the results obtained from the dual variables approach~\cite{Gattringer:2012df}. Our results agree very well which strongly
supports the conjecture that the tangent plane is equivalent with the original integration domain. We note
that we observed neither instabilities nor runaways in our simulations, in contrast with the experiences 
reported by Cristoforetti~{\it et~al}~\cite{Cristoforetti:2013wha}, even when we set $h$ to purely real values,
as they did in their study.

%The particularly fortuitous situation that the tangent space $T_1$ of the global minimum thimble $\mathcal{T}_1$ is in the same homology class as the original integral over $\mathds{R}^N$ has been explicitly shown to occur in fermionic models~\cite{Alexandru:2015sua}. When this occurs, an integration over $T_1$ will yield the same result as an integration over $\mathds{R}^N$, and consequently is equivalent to an integral over all thimbles. In many cases, the sign problem is reduced enough on $T_1$ that observables can simply be reweighted as in~\ref{reweighting} (see \ref{real_problem} below). It should be noted that the sign problem scales exponentially with $\beta V$, so as the spacetime volume grows, the option of reweighting an integration over $T_1$ fades at some point. At such spacetime volumes, flowing the plane is necessary to tame the sign problem. Nevertheless, it is advantageous to integrate over $T_1$ whenever possible for several reasons.  First, such integrations are numerically very cheap as they do not require the flow. Moreover, in these cases it is only necessary to find the global minimum critical point. This is a great benefit to field theoretical models (in particular) because field theoretical models have many critical points, most of which are difficult to find.

\begin{figure}[b]
\centering
\includegraphics[width=\columnwidth]{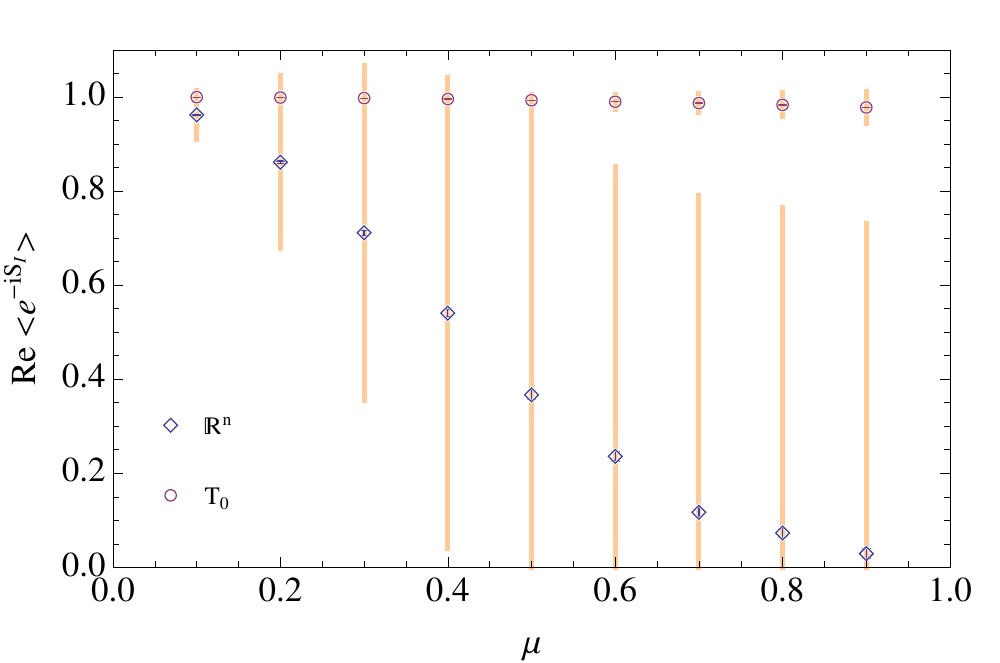}
\caption{Dependence of the average phase on the chemical potential $\mu$ below the critical transition on the 
real plane and the tangent plane $T_0$ for a $4^4$ lattice with the parameters $m=\lambda=1.0$. The orange bars 
indicate the standard deviation of $\Re e^{-iS_I}$. The phase hardly varies for these parameters, so in this case 
it is sufficient to shift the integration domain to $T_0$ to tame the sign problem.\label{fig:fluctuations}}
\end{figure} 

The simulations above are feasible even at $T_\text{flow}=0$ because the sign fluctuations are significantly smaller
when integrating over the tangent space rather than the original integration domain. In Fig.~\ref{fig:fluctuations}
we show the average phase, $\text{Re}\av{\text{e}^{-i S_I}}$, and its standard deviation as a function of the chemical potential in
the symmetric phase, both for simulations using $\mathds{R}^n$ as a starting manifold and $T_0$. We 
show in the graph both the standard deviation and the standard deviation of the mean of $\text{Re}\av{\text{e}^{-i S_I}}$
 so that the number of sampled points in the simulations 
do not play a role in the comparison. We see that as we approach the transition point, the phase fluctuates rapidly
when we sample points on $\mathds{R}^n$ and the number of sampled configurations required to measure the observables
with a given precision grows very quickly. In contrast, the phase fluctuations on $T_0$ are very mild, at least for
the lattice volume used in these simulations, and the phase can be easily accounted for by reweighting.

%This situation was first postulated to occur in the relativistic bose gas in~\cite{Cristoforetti:2013wha}. An explicit verification that $T_1$ is in the same homology class as $\mathds{R}^N$ is very difficult, however, because it requires a characterization of the asymptotics of the real part of the action $S_R$ in a high dimensional complex space. Despite the lack of a proof, there is compelling numerical evidence to believe that an integral over $T_1$ is equivalent to an integral over $\mathds{R}^N$. First, at no point in a Monte Carlo integration do any ``runaways" occur on $T_1$ in our implementation, which indicates that the action is likely bounded from below on the tangent space.  Furthermore, integrations over $T_1$ yield results consistent with results from studies using complex Langevin methods~\cite{Aarts:2008wh} and dual variables~\cite{Gattringer:2012df}. Given these two facts, we proceed under the (reasonable) assumption that $T_1$ is in the same homology class as $\mathds{R}^N$. It should be noted that runaways have been noted to occur \cite{Cristoforetti:2013wha}, and their cause is not obvious. Naively a Stokes line might cause a runaway due to the presence of flat directions, indeed \cite{Cristoforetti:2013wha} use a real symmetry breaking, resulting in a Stokes line between $\phi_+$ and $\phi_-$, however the reality of $h$ is very likely not the cause of the runaways, because we have seen no runaways in any Monte Carlo in our implementation with a purely real symmetry breaking.

\begin{figure*}[t]
\includegraphics[width=0.48\textwidth]{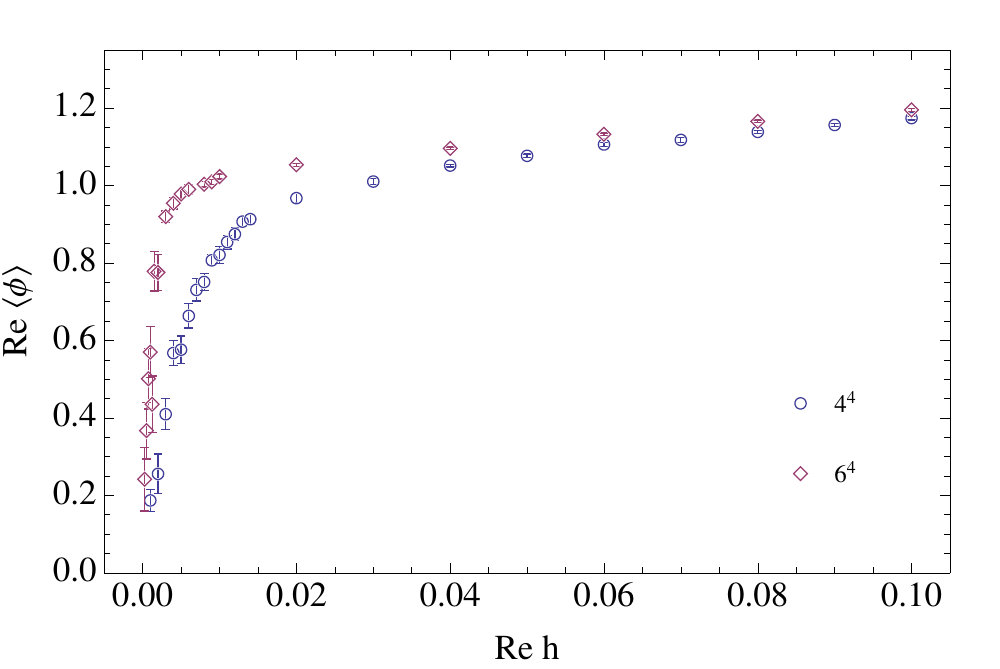}\kern1cm
\includegraphics[width=0.48\textwidth]{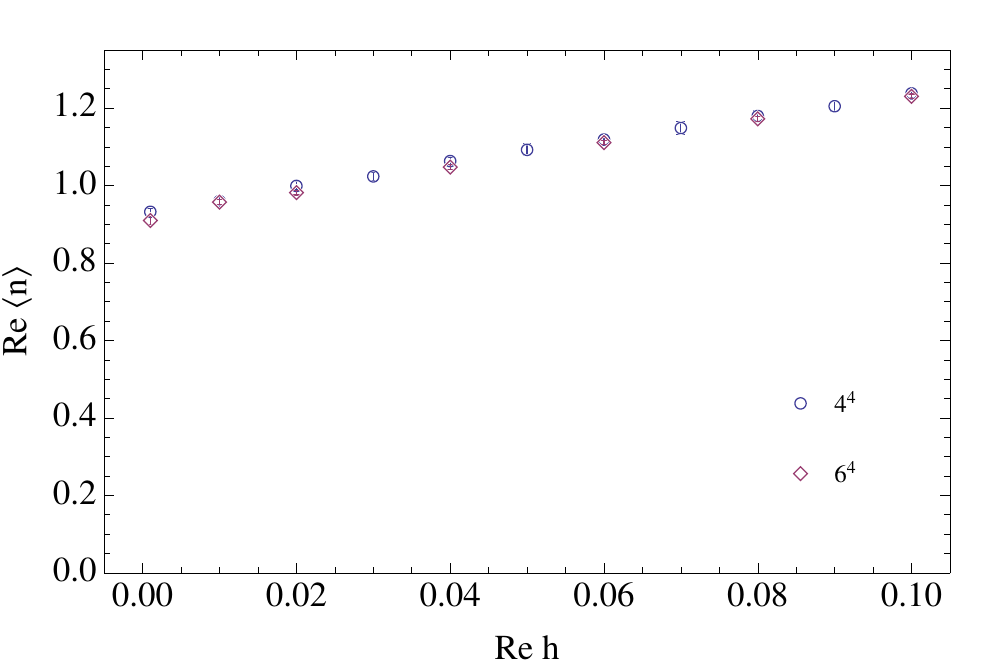}
\caption{Observables as a function of the symmetry breaking parameter $h$ on a $4^4$ lattice (blue) and $6^4$ lattice (red) with $m=\lambda=1.0$, $\mu=1.3$. 
Note that the charge varies slowly as we decrease $h$, whereas for $\Re h\lesssim 0.02$ the order parameter $\av{\phi}$ becomes
dominated by finite volume effects and, as expected, approaches zero as $h\to0$. As expected, the values of $h$ for which there is a significant finite volume effect decreases as the volume is increased.
\label{fig:obs_vs_h}}
\end{figure*}

As discussed earlier, when the symmetry breaking parameter $h$ goes to zero, the order parameter
\beq
\av{\phi} = \frac1{V_4} \pd{\log Z}h \,,
\eeq
vanishes if the limit $h\rightarrow 0$ is taken while the volume is kept fixed. In the symmetry broken phase, this is a finite volume effect. 
This happens because when $h$ is small the fluctuations in $\phi$ are large enough to overcome the potential
barrier between $\phi_+$ and $\phi_-$, $\Delta S = V_4 4h \sqrt{(\hat\mu^2-m^2)/(2\lambda)}+{\cal O}(h^2)$.
Since we discuss here simulations at fixed volume, it is important to determine the range of $h$ where
these finite volume effects are important. In Fig.~\ref{fig:obs_vs_h} we plot the charge and the order
parameter $\av{\phi}$ as a function of $h$, as determined from simulations over $T_+$. We set the value of
the chemical potential to $\mu=1.3$, to make sure we are in the high-density, symmetry broken phase. 
In these simulations the ratio between the imaginary and real part of h is kept fixed: $\Re h/ \Im h = 10$.
We see that for values smaller than $\Re h\lesssim 0.02$ (for the $4^4$ lattice) and  $\Re h\lesssim 0.01$ (for the $6^4$ lattice) the finite volume effects become important and restore the (near) symmetry.
For the following calculation we use a value of $h=0.1\times(1+i/10)$ to make sure we are away from the
region where finite volume effects dominate. 

Note that for these simulations, at small values of $h$, there is a nearly flat direction in field space that is sampled inefficiently
using the proposals discussed earlier. This can be easily fixed. The flat direction corresponds roughly
to a circle in the two-dimensional space spanned by $\hat\rho_0$, the ``eigenvector'' nearly parallel with the constant
field with $\phi_1=-\phi_2$ (the ``Goldstone" direction) and ``eigenvector'' $\hat\rho_1$ nearly parallel with the constant field direction with $\phi_1=\phi_2$ (the ``Higgs" direction).
The updates along the other directions are treated as discussed above, but in the $\hat\rho_{0,1}$ plane we use a polar representation 
and update the angular part with steps of size $\Delta/\sqrt{\lambda_0}$ and the radial part with steps of size $\Delta/\sqrt{\lambda_1}$.
The polar coordinates are defined in relation to the center $c_0=0$ and 
$c_1=(\phi_--\phi_+)\cdot\hat\rho_1/2$, with $c_{0,1}$ the coordinates in the $\hat\rho_{0,1}$ basis.
To preserve detailed balance we have to modify the accept/reject step to take into account the asymmetry in this polar proposal,
that is we accept the update with probability $\exp[-S_R(\phi')+S_R(\phi)] r'/r$, where $r$ and $r'$ represent the radial
coordinates for $\phi$ and $\phi'$ in the polar representation.

%Given the fact that the goal is to analyze SSB in this system, it is first necessary to understand at what value of the symmetry breaking $h$ the system is in a universal regime. The transition in the order parameter $\langle \phi \rangle$ occurs around $h\sim 0.01(1+i/10)$. Analyzing this system in its linear regime yields behavior consistent with SSB, therefore we choose $h=0.1(1+i/10)$, well within the linear region, in subsequent analysis.

%\subsection{Single thimble}
%Argue that's a ``one thimble" calculation. $T_{flow}\rightarrow\infty$ versus metropolis steps $\rightarrow\infty$.
%Argue that $\langle greg \rangle$ improves.
%Agreement with $T_{flow}=0$. No evidence for sizable contributions from other thimbles in this regime.

As we discussed in Section~\ref{sec:phi4}, to obtain exact results for this model using thimble sampling 
we have to consider a set of thimbles, at the very least ${\cal T}_+$ and ${\cal T}_-$. One of the questions 
we want to address here is whether the single thimble calculation is sufficient to recover the exact result. 
Note that previous calculations for the relativistic boson gas that used Lefschetz thimbles were either carried out in 
the tangent plane~\cite{Cristoforetti:2013wha}, which as we argued provides the exact result, or 
without including the symmetry breaking term in which case the critical point is actually a circle and the integration 
was done over the entire set of thimbles attached to this circle~\cite{Fujii:2013sra}. As such, this question 
has not yet been directly addressed.

In Section~\ref{sec:algo} we showed that to perform a one thimble calculation with the contraction algorithm, 
say ${\cal T}_\sigma$, we need to sample the manifold produced by flowing the tangent space
of the thimble at the critical point $T_\sigma$, in the limit $T_\text{flow}\to\infty$. If we start at
the critical point $\phi_\sigma$ and perform only small step updates, for large flow times the
algorithm will only sample ${\cal T}_\sigma$. We focus here on the symmetry broken phase and
we carry out a simulation for $\mu=1.3$. The thimble we sample is ${\cal T}_+$. 
As the limit $T_\text{flow} \to \infty$ cannot be taken in practice, it is necessary to 
devise an operational definition. We carry out simulations at increasing flow times and monitor
the observables of interest and the imaginary part of the action on the sampled manifold. As the
sampled manifold approaches the thimble the observables should converge to their average over
the thimble and the fluctuations of the imaginary part of the action around $S_I(\phi_+)$ should
be reduced drastically in amplitude (in the limit $T_\text{flow}\to\infty$ all points in the sampled pocket
around $\phi_+$ should have the same $S_I$).

%one samples the tangent space of the thimble according to the effective action $S_\text{eff}$, then takes the limit $T_{flow} \rightarrow \infty$ while keeping the number of Metropolis $\mathcal{N}$ steps fixed. Then to obtain a one thimble result, in principle, the limit $\mathcal{N} \rightarrow \infty$ is taken. We choose $\mathcal{N}=5\times 10^6$ for our simulations, and perform one thimble calculations at $\mu=1.3$ and $\mu=1.7$ in the spontaneously broken region of $h=0.1(1+i/10)$. As the limits $T_\text{flow} \to \infty$ and $\mathcal{N} \to \infty$ cannot actually be taken in practice, it is necessary to devise an operational definition of a one thimble calculation in a finite Monte Carlo. Two criteria are chosen here: the fluctuations in $e^{-iS_I}$ must become small and the observable in question must asymptote to a fixed value in the flow. The first criterion is chosen simply because the thimble is a surface of constant $S_I$, so a calculations converging to an integration over one thimble must have small fluctuations in $S_I$. The second criterion reflects the fact, in order to accurately calculate observables, the entire relevant region of the thimble (that is, the portion of the thimble where the probability density $e^{-S_\text{eff}}/Z$ is appreciable) must explored be by the Markov Chain.

\begin{figure*}[t]
\includegraphics[width=\textwidth]{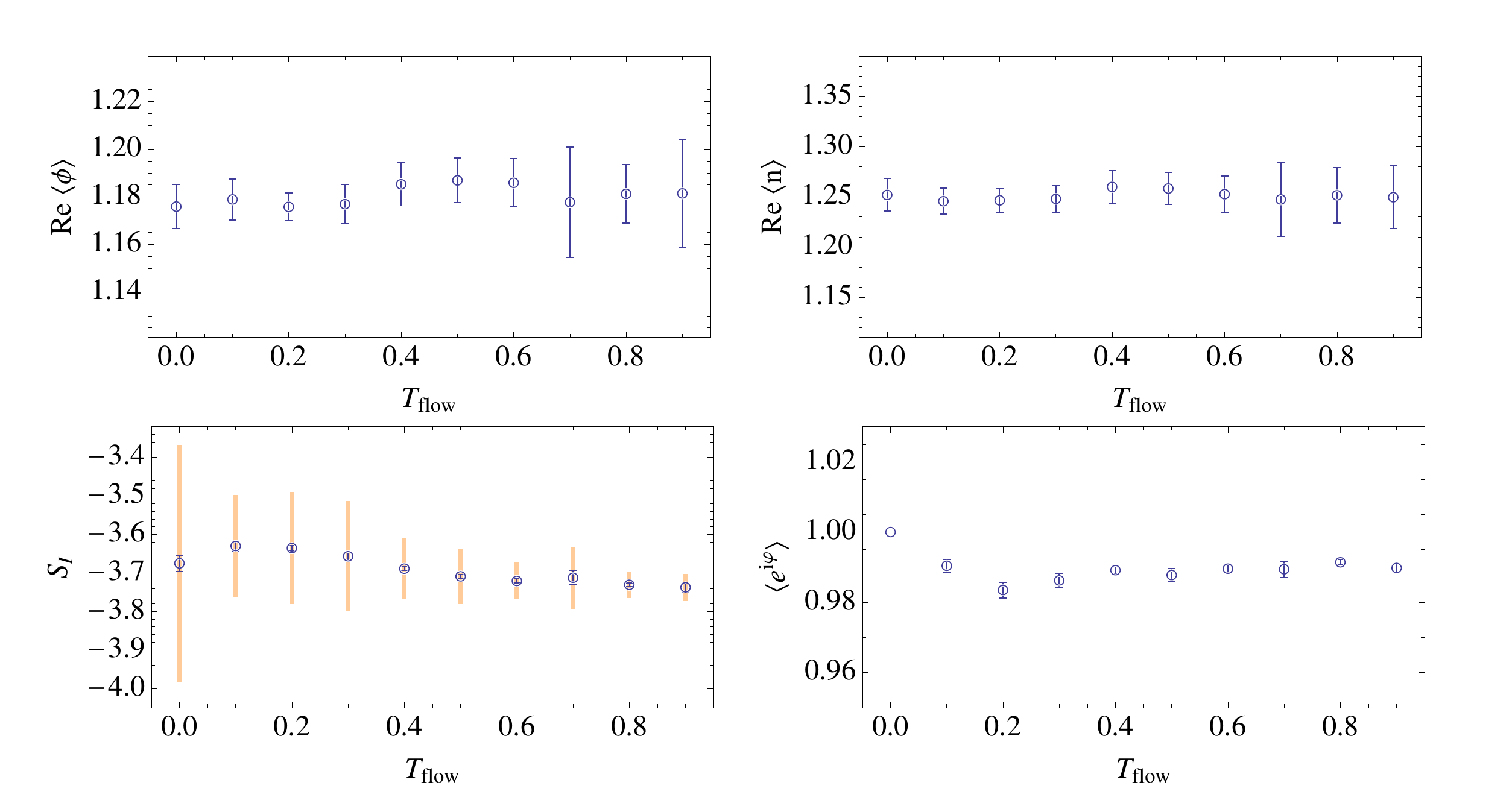}
%\caption{$T_\text{flow}$ extrapolations for several observables on a $4^4$ lattice with parameters $m=\lambda=1.0$, $\mu=1.3$, $h=0.1(1+i/10)$.}
%\includegraphics[width=\textwidth]{figs_round2/obs_vs_tflow_mu1p7}
%\caption{$T_\text{flow}$ extrapolations for the charge and the order parameter on a $4^4$ lattice with 
%$m=\lambda=1.0$, $h=0.1(1+i/10)$. Two values of the chemical potential are used: $\mu=1.3$ (top) and 
%$\mu=1.7$ (bottom).\label{fig:higher_mu}}
\caption{$T_\text{flow}$ extrapolations for the order parameter and the charge~(top row) and the average imaginary 
part of the action $S_I(\phi_f)$ 
and the residual phase $\exp i\varphi=d\phi/|d\phi|$ which is computed as $\varphi=\Im \log \det J$~(bottom row) on a $4^4$ lattice with 
$m=\lambda=1.0$, $h=0.1(1+i/10)$, and $\mu=1.3$. For $S_I$ we also indicate the standard deviation to gauge
its fluctuations.\label{fig:onethimble}}
\end{figure*} 

We carried out simulations for $\mu=1.3$, $h=0.1\times(1+i/10)$ and measured the order parameter and 
the charge as a function of $T_\text{flow}$. For each simulation we carried out $5\times 10^6$ updates 
and we adjusted the step size to get an acceptance rate close to 50\%. Note that the step size needs 
to be decreased dramatically, since the updates are carried out in the parametrization manifold, where 
the distribution of configurations become tightly packed around the critical point. We have to use 
anisotropic proposals along the ``eigenvectors'' in the tangent plane to take into account the fact 
that the flow compresses these directions differently, which is done by adjusting the step size in 
direction $\hat\rho_i$ to $\Delta \exp(-\lambda_i T_\text{flow})/\sqrt{\lambda_i}$~\cite{Alexandru:2015xva}. 
The measurements were carried out on configurations separated by $10,000$ updates.

In Fig.~\ref{fig:onethimble} we show our results. We see that the values of $\av{\phi}$ and $\av{n}$ 
show very little dependence on the flow time, indicating that the single thimble result is equal to the 
exact result, at least at the level of the errobars. In the bottom row of the figure we also plot the 
value of the imaginary part of the action for the configurations sampled in our Monte Carlo simulations. 
It is clear that the fluctuations die out quite quickly, and as $T_\text{flow}$ approaches $1.0$ we 
have almost no fluctuations, indicating that we are sampling ${\cal T}_+$ only. In the bottom left panel
of Fig.~\ref{fig:onethimble} we plot the residual phase, that is the phase related to the curvature 
of the sampled manifold, as a function of flow time. In our formulation, this is the part of the
reweighting phase that is due to the imaginary part of the Jacobian, that is $\Im\log\det J$. In
the limit $T_\text{flow}\to\infty$ the fluctuations of $S_I$ vanish, but the residual phase continues
to fluctuate. These fluctuations are not expected to lead to a sign problem, since the thimbles 
are expected to be relatively smooth at least in the region where the bulk of the contribution to
the integral comes from. This is indeed what we find: the phase fluctuations are very close to one
for all intermediate manifolds and asymptote to $\av{\exp(i\Im\log\det J)}=0.989(1)$. The closeness
of this average to $1$ also indicates that the thimble is quite flat. Incidentally,
our result for the average residual phase is in agreement with the one determined by Fujii~{\it et al}~\cite{Fujii:2013sra}
for $\mu=1.3$ using the HMC algorithm: $\av{\exp(i\Im\log\det J)}=0.99(3)$. Note that they sample a
different manifold than ${\cal T}_+$ ($h=0$ in that calculation), so it is not clear that a geometric measure like the average
residual phase should be identical.

%end comment

We conclude that in the regime where the finite volume effects are small,
the partition function is dominated by the contribution of one thimble ${\cal T}_+$. 
However, it is still worthwhile investigating the contribution due to other thimbles
for two reasons. First, if we are interested in studying the system when the 
symmetry breaking term is small and the finite volume effects are important, 
it is likely that we need to include the contribution from other thimbles. Secondly,
while we showed that the two observables we measured seem to be saturated by the contribution 
of ${\cal T}_+$, it is possible that these observables are special in the sense of being insensitive to removing the samples due to other
thimbles. 

Semiclassical arguments suggest the contribution of subdominant thimbles vanishes.
If we consider the ratio of the two contributions to the partition function
\beq
\frac{Z_+}{Z_-} \equiv \frac{\int_{{\cal T}_+} |d\phi|\, e^{-S_R(\phi)}}{\int_{{\cal T}_-} |d\phi|\, e^{-S_R(\phi)}}
\eeq
we expect that it goes to infinity very quickly as $\Re h$ increases. 
At leading order in the semiclassical expansion this ratio is
\beq
\frac{Z_+}{Z_-} \approx e^{-[S_R(\phi_+)-S_R(\phi_-)]} \approx e^{4{\Re h}V_4\sqrt{(\hat\mu^2-m^2)/(2\lambda)}} \,,
\eeq
which justifies our expectation. The next-to-leading order estimate
includes the gaussian fluctuations 
\beq
\frac{Z_+}{Z_-} \approx e^{-[S_R(\phi_+)-S_R(\phi_-)]} 
\sqrt{\frac{\widetilde\det H(\phi_-)}{\widetilde\det H(\phi_+)}} \,,
\eeq
with $H(\phi)$ the hessian at $\phi$ and $\widetilde\det H(\phi)$ being the product of 
positive ``eigenvalues'' $\lambda_i$ as defined in Eq.~\ref{eq:heigensys}.
The next to leading order ratio increases more slowly than the leading order
estimate and it is possible that 
the semiclassical arguments could break down. In the remainder of this section we will study numerically 
the relative contribution due to ${\cal T}_-$ and compare it with the semiclassical approximation.

We present here a method to sample two thimbles in the context of the contraction algorithm.
Assuming that the original integration domain decomposes into a sum over ${\cal T}_+$ and
${\cal T}_-$, the average observable is given by
\beq
\av{\cal O} = \frac{n_+\int_{{\cal T}_+} d\phi\,e^{-S(\phi)}{\cal O}(\phi) + n_-\int_{{\cal T}_-} d\phi\,e^{-S(\phi)}{\cal O}(\phi)}
{n_+\int_{{\cal T}_+} d\phi\,e^{-S(\phi)} + n_-\int_{{\cal T}_-} d\phi\,e^{-S(\phi)}} \,.
\eeq
Notice that both thimbles appear in both the numerator and the denominator, which makes a decomposition of 
$\av{\cal O}$ into a straightforward sum of integrals over thimbles not possible. The integers $n_+$ and
$n_-$ count the number of intersection points between the parametrization manifold and the unstable thimbles ${\cal K}_+$
and ${\cal K}_-$. Each of them lies within a different pocket in the parametrization manifold. In our simulations on the tangent plane $T_+$
we have only seen evidence for one intersection point with ${\cal K}_+$ at $\phi_+$ and one intersection
point close to $\phi_-$~($\phi_-$ is not included in  $T_+$ but it is nearby), which we assume is associated with ${\cal K}_-$. Here we will discuss a method to sample only these two pockets,
 implicitly assuming that $|n_\pm|=1$; the sign of $n_\pm$ is automatically taken into account 
correctly by the flow.

To sample ${\cal T}_+$ and ${\cal T}_-$ simultaneously, we use
the contraction algorithm with large $T_\text{flow}$ and set $T_+$ as the parametrization manifold, which we
assume to be equivalent with $\mathds{R}^n$. The only difference is that instead of starting close to $\phi_+$
and making only small proposals, we interweave these updates with large proposals that move us from the vicinity of
$\phi_+$ to the neighborhood of $\phi_-$. Since we parametrize the configurations in $T_+$ using 
coefficients in the $\hat\rho_i$-basis, we implement this proposal by flipping the sign of the coefficient
corresponding to the eigenvector $\hat\rho_1$, that is nearly parallel with the constant field $\phi_1=\phi_2$ configuration.
In the limit that $T_\text{flow}\to\infty$ this process will sample the two thimbles according to the probability
density
\beq\label{joint_dist}
\frac{d\text{Pr}_\pm(\phi)}{|d\phi|}= e^{-S_{R}(\phi)} \Big/\int_{{\cal T}_+ \cup {\cal T}_-}|d\phi|\,e^{-S_R(\phi)} \,.
\eeq
As before, the residual phase needs to be folded in the observable in the reweighting step. Additionally, there is
another fluctuating phase, even in the $T_\text{flow}\to\infty$ limit, since the imaginary 
part of the action is different on the two thimbles ~($S_I(\phi_+)\not=S_I(\phi_-)$). In the regime we study here, both these phases 
lead to mild fluctuation and there is no problem reweighting them.

\begin{figure}[t]
\centering
\includegraphics[width=\columnwidth]{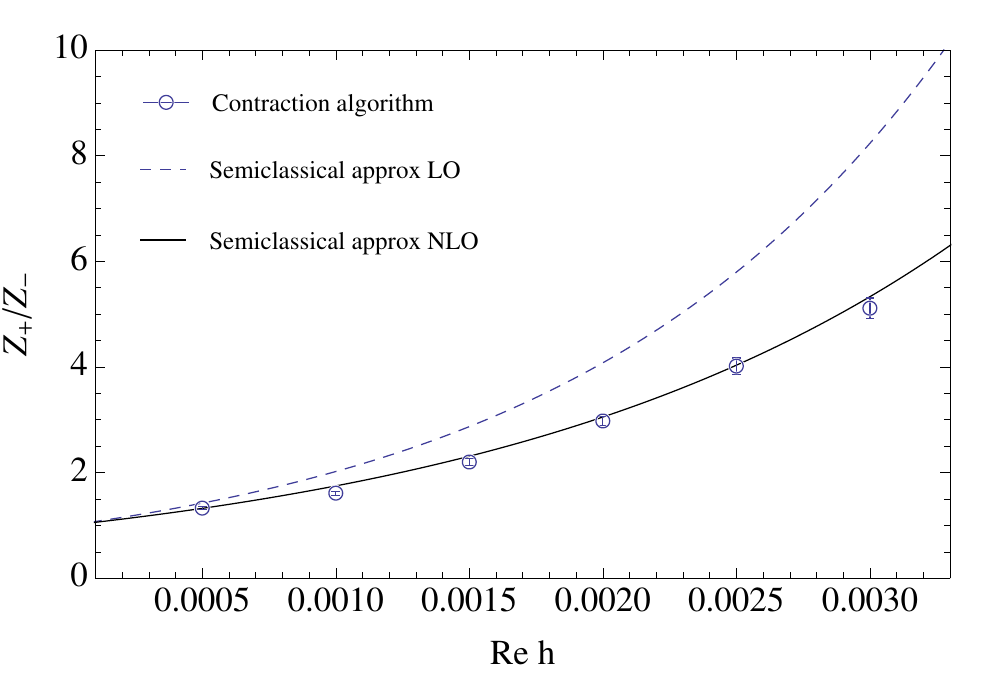}
\caption{
The relative weights of thimble contribution as determined from sampling with the contraction algorithm,
compared to the predictions from semiclassical approximation.
\label{fig:zpzm}}
\end{figure} 

\begin{table}[b]
\begin{tabular*}{0.99\columnwidth}{@{\extracolsep{\stretch{1}}}*{4}{r}@{}}
\toprule
$h_R$ &  $Z_1/Z_2$ & LO & NLO  \\
\midrule
$5\times 10^{-4}$  & $1.33(3)$ & 1.42 & 1.32\\
$10\times 10^{-4}$ & $1.61(4)$ & 2.02 & 1.75\\
$15\times 10^{-4}$ & $2.20(6)$ & 2.87 & 2.31\\
$20\times 10^{-4}$ & $2.98(8)$ & 4.08 & 3.05\\
$25\times 10^{-4}$ & $4.02(15)$ & 5.79& 4.03 \\
$30\times 10^{-4}$ & $5.11(20)$ & 8.23& 5.33 \\
$0.02$   & $10^7/0$ & $1.27 \times10^6$ & $6.96 \times 10^5$\\
$0.1$     & $10^7/0$ &  $2.67 \times 10^{30}$ &$1.01\times 10^{24}$\\
\bottomrule
\end{tabular*}
\caption{Distribution of points among the two lowest action thimbles $\mathcal{T}_+$ and $\mathcal{T}_-$ over 
the course of a $10 \times 10^6$ step Monte Carlo on a $4^4$ lattice with parameters $m=\lambda=1.0$, 
$\mu=1.3$, $T_\text{flow}=1.0$. For large values of $h$ not a single transition occurs between 
$\mathcal{T}_+$ and $\mathcal{T}_-$ over the entire course of a Monte Carlo simulation. The results are
compared with the leading order (LO) and next to leading order (NLO) semiclassical predictions.
\label{tab:zpzm}}
\end{table}

Note that for computing the reweighting factor we do not need to identify the thimble associated with 
any of the sampled points. The flow automatically
produces the right result. In fact, the identification might not even be possible for
small $T_\text{flow}$ since the separation into ${\cal T}_\pm$ contributions is not sharply defined.
However, as we increase $T_\text{flow}$ it is easy to identify the associated thimble for
each of the configuration sampled in $T_+$ since they concentrate very sharply around $\phi_\pm$.
We run a set of $10\times10^6$ updates for $\mu=1.3$, $m=1$, $\lambda=1$ and a series
of values for $h$, while we keep as before $\Re h/\Im h = 10$. We set $T_\text{flow}=1$, so that
the associated thimble can be identified easily and the ratio of sampled points in
${\cal T}_+$ and ${\cal T}_-$ can be computed. The results are plotted in Fig.~\ref{fig:zpzm}. We see that
for small values of $h$, where ${\cal T}_-$ has a non-negligible contribution, this ratio is
actually very close to the  predictions of the next-to-leading order semiclassical approximation.
We conclude that the non-gaussian fluctuations do not affect the sampling ratio significantly
and that the semiclassical arguments can be trusted. In Table~\ref{tab:zpzm} we record the
measured values for this ratio. We also include the results for two simulations with large values of
$h$, similar to the ones we discussed earlier in this section, and we see that the subdominant
thimble ${\cal T}_-$ is never visited, in agreement with semiclassical prediction.
We conclude that for values of $h$ where the finite volume effects are small, the subdominant
thimbles are indeed unlikely to contribute significantly.

%\begin{table}[b]
%\centering
%\begin{tabular}{|| c | c | c | c | c ||}
%\hline
%h & $\# \mathcal{T}_1$ & $\# \mathcal{T}_2$ & $\text{ratio}$ & $\text{e}^{S_R(\phi_-)-S_R(\phi_+)}$  \\
%\hline \hline
%5 \times 10^{-4}
%0.003 & 7871 & 1984 & 3.97 & 8.21 \\
%0.02 & 10000 & 0 & 0 & $1.28 \times10^6$ \\
%0.1 & 10000 & 0 & 0 &  $4.58 \times 10^{30}$\\
%\hline
%\end{tabular}

%% file: conclusions.tex
\section{Conclusions}
\label{sec:conclusions}

We analyzed the relativistic Bose gas using contraction algorithm, presenting the first application of the contraction algorithm to a quantum field theory. In studying the high density broken phase of the theory, we noticed the existence of a Stokes line and fixed this problem by using a complex value of the symmetry breaking parameter $h$. We verified that the results obtained for the charge density agreed with previous calculations when available. 
We then focused on the order parameter $\av{\phi}$, which is sensitive to spontaneous symmetry breaking. We first determined the values of $h$ for which finite volume effects, which tend to restore the symmetry, are small. 
The results from complex Langevin calculations agree with our tangent plane calculations, lending support to the conjecture that the tangent plane is indeed equivalent to the real plane. In contrast to  \cite{Cristoforetti:2013wha} we found no runaway trajectories, as expected if the tangent plane is in fact equivalent to the real plane.

We then showed how to use the  contraction algorithm to isolate the contribution of a single thimble.
The results obtained from the single thimble agree with the results obtained from the tangent plane within a few percent, indicating that the contributions from other thimbles are negligible for the parameters explored.
We generalized the contraction algorithm to perform calculations
over two thimbles. 
 We found that the contribution from the sub-dominant thimble follows closely the semiclassical estimates (negligible at large $h$ but comparable to the leading one at smaller~$h$).
 
 Our calculations point towards some obvious extensions and generalizations. The most pressing one is perhaps the extension to the thermodynamic limit, with the goal of determining whether the contribution of any other thimble survives this  limit.
 Also, our group has recently developed similar technology to that described in this paper to study the real time dynamics of a simple quantum mechanical model \cite{Alexandru:2016gsd}. We look forward to combining the experience acquired in the present paper to study the real time dynamics of the Bose gas with an eye towards the computation of transport coefficients.